\documentclass[12pt]{article}
\usepackage{amssymb,amsmath,epsfig}

\begin{document}

\title{\bf Dynamics of Viscous Dissipative Plane Symmetric
Gravitational Collapse}
\author{M. Sharif \thanks {msharif@math.pu.edu.pk} and Zakia
Rehmat
\thanks {zakiarehmat@gmail.com}\\
Department of Mathematics, University of the Punjab,\\
Quaid-e-Azam Campus, Lahore-54590, Pakistan.}

\date{}

\maketitle
\begin{abstract}
We present dynamical description of gravitational collapse in view
of Misner and Sharp's formalism. Matter under consideration is a
complicated fluid consistent with plane symmetry which we assume to
undergo dissipation in the form of heat flow, radiation, shear and
bulk viscosity. Junction conditions are studied for a general
spacetime in the interior and Vaidya spacetime in the exterior
regions. Dynamical equations are obtained and coupled with causal
transport equations derived in context of M$\ddot{u}$ller Israel
Stewart theory. The role of dissipative quantities over collapse is
investigated.
\end{abstract}


\section{Introduction}

The ultimate fate of the star (when it undergoes catastrophic phase
of collapse) is one of the most important questions in gravitation
theory today. When a star has exhausted all of its nuclear fuel, it
collapses under the influence of its own gravity and releases large
amount of energy. In fact, it is a highly dissipative process, i.e.,
energy is not conserved in it, rather due to various forces and with
the passage of time, it becomes lesser. Dissipative process plays
dominant role in the formation and evolution of stars.

The initial discussion over this problem was given by Oppenheimer
and Snyder \cite{1} who assumed a spherically symmetric distribution
of matter. They took the most simplest form of matter, i.e., dust
and the flow is considered to be adiabatic. It is somewhat
unrealistic to ignore the pressure as it cannot be overlooked in the
formation of singularity. Misner and Sharp \cite{2} adopted a better
approach by considering an ideal fluid which gave a more realistic
analysis of gravitational collapse. Both of them assumed vacuum in
the exterior region. Vaidya \cite{3} introduced a non-vacuum
exterior by giving the idea of outgoing radiation in collapse. It
was physically a quite reasonable assumption as radiation is a
confirmation that dissipative processes are occurring, causing loss
of thermal energy of the system which is an effective way of
decreasing internal pressure.

The Darmois junction conditions  \cite{4} gave a way to obtain
exact models of an interior spacetime with heat flux to match with
exterior Vaidya spacetime. Sharif and Ahmad \cite{5} considered
the perfect fluid with positive cosmological constant to discuss
the junction conditions with spherical symmetry. The same authors
\cite{6} also worked on junction conditions for plane symmetric
spacetimes.

Goswami \cite{7} made an attempt in search of a more physical
model of collapse. He considered dust like matter with heat flux
to conclude that dissipation causes a bounce in collapse before
the formation of singularity. Nath at el. \cite{8} investigated
dissipation in the form of heat flow and formulated junction
conditions between charged Vaidya spacetime in exterior and
quasi-spherical Szekeres spacetime in interior regions. They also
discussed apparent horizons and singularity formation. Ghosh and
Deshkar \cite{9} studied gravitational collapse of radiating star
with plane symmetry and pointed out some useful results. A lot of
work is being done over gravitational collapse by considering
shear free motion of the fluid. Although, it leads to
simplification in obtaining exact solutions of the field
equations, yet it is an unrealistic approach. Shear viscosity is a
source of dissipating energy and plays an important role in
collapse. Chan \cite{10} investigated gravitational collapse, with
radial heat flow, radiation and shear viscosity. He showed how the
pressure became anisotropic due to shear viscosity.

Herrera and Santos \cite{11} discussed the dynamics of
gravitational collapse which undergoes dissipation in the form of
heat flow and radiation. Di Prisco et al. \cite{12} extended this
work by adding charge and dissipation in the form of shear
viscosity. Herrera \cite{13} provided comprehensive details of
inertia of heat and how it plays an effective role in dynamics of
dissipative collapse. Herrera and Martinez \cite{14} presented
relativistic model of heat conducting collapsing object and
debated over the effect of a parameter which occurs in dynamical
equation on collapse. Herrera and collaborators
\cite{15}-\cite{16} proposed a model of shear free conformally
flat collapse and focused on the role of relaxation process, local
anisotropy and relation between dissipation and density
inhomogeneity.

Recently, Herrera et al. \cite{17} threw light on behavior of
non-equilibrium massive object which lost energy due to heat flow,
radiation, shear and bulk viscosity. Matter under consideration was
distributed with spherical symmetry. It has become quite clear that
when mass and energy densities involved in the physical phenomenon
are sufficiently high as in gravitational collapse, gravitational
field plays an important and dominant role. The gravitational
dynamics then must be taken into account for a meaningful
description of such ultra high energy objects. This fact motivated
us to elaborate the above mentioned paper in the context of plane
symmetries. Matter under consideration is a complicated fluid which
suffers through dissipation. Misner and Sharp's prescription is used
to work out dynamical equations. Transport equations are obtained in
the context of M$\ddot{u}$ller Israel Stewart theory \cite{18},
\cite{19} which is a causal theory for dissipative fluids.
Thermodynamic viscous/heat coupling coefficients are taken to be
non-vanishing which is expected to be quite plausible in non-uniform
stellar models of universe. One of the dynamical equations is then
coupled to transport equations in order to figure out the influence
of dissipation over collapse.

The paper is written in the following manner. The next section is
about the matter distribution in the interior region and some
physical quantities relevant to matter under consideration. The
Einstein field equation are worked out in section \textbf{3} and
junction conditions are discussed in section \textbf{4}. Dynamical
equations are formulated in section \textbf{5} and are coupled to
transport equations in section \textbf{6}. The last section
discusses and concludes the main results of the paper.

\section{Interior matter distribution and some physical quantities}

A $4$-dimensional spacetime is split into two regions: interior
$V^{-}$ and exterior $V^{+}$ through a hypersurface $\Sigma$ which
is the boundary of both regions. We assume the matter distribution
in the interior region to be consistent with plane symmetry. The
interior region $V^{-}$ admits the following line element
\begin{equation}\label{1}
ds^{2}_{-}=-f(t,z)dt^{2}+g(t,z)(dx^{2}+dy^{2})+h(t,z)dz^{2},
\end{equation}
where ${\{\chi^{-\mu}\}}\equiv\{t,x,y,z\}~(\mu=0,1,2,3)$. The fluid
is presumed to dissipate energy in terms of heat flow, radiation,
shearing and bulk viscosity.

The energy-momentum tensor for such a fluid is defined as
\begin{equation}\label{2}
T_{ab}=(\mu+p+\Pi)V_{a}V_{b}+(p+\Pi)g_{ab}+q_{a}V_{b}+q_{b}V_{a}+\epsilon
l_{a}l_{b}+\pi_{ab},
\end{equation}
where $\mu,~p,~\Pi,~q_a,~l_a$ and $\pi_{ab}$ are the energy
density, pressure, bulk viscosity, heat flow, null four-vector in
$z$-direction and shear viscosity tensor respectively. Heat flow
$q_{a}$ is taken to be orthogonal to velocity $V^{a}$, i.e.,
$q_{a}V^{a}=0$. Moreover, we have
\begin{eqnarray}\label{3}
V^{a}V_{a}=-1,\quad l^{a}V_{a}=-1,\quad\pi_{ab}V^{b}=0, \quad
\pi_{[ab]}=0,\quad\pi^{a}_{a}=0,\quad l^{a}l_{a}=0.
\end{eqnarray}
In the standard irreversible thermodynamics by Eckart, we have the
following relation \cite{20}
\begin{equation}\label{4}
\pi_{ab}=-2\eta\sigma_{ab},\quad \Pi=-\zeta\Theta,
\end{equation}
where $\eta$ and $\zeta$ stand for coefficients of shear and bulk
viscosity, $\sigma_{ab}$ is the shear tensor and $\Theta$ is the
expansion. The algebraic nature of Eckart constitutive equations
causes several problems but we are concerned with the causal
approach of dissipative variables. Thus we would not assume
(\ref{4}) rather we shall resort to transport equations of
M$\ddot{u}$ller-Israel-Stewart theory.

The shear tensor $\sigma_{ab}$ is defined as
\begin{equation}\label{5}
\sigma_{ab}=V_{(a;b)}+a_{(a}V_{b)}-\frac{1}{3}\Theta h_{ab},
\end{equation}
where the acceleration $a_{a}$ and the expansion $\Theta$ are
given by
\begin{equation}\label{6}
a_{a}=V_{a;b}V^{b},\quad \Theta=V^{a}_{;a}
\end{equation}
and $h_{ab}=g_{ab}+V_{a}V_{b}$ is the projection tensor. The shear
tensor $\sigma_{ab}$ satisfies
\begin{equation}\label{7}
V_{a}\sigma^{ab}=0,\quad\sigma^{ab}=\sigma^{ba},\quad\sigma^{a}_{a}=0.
\end{equation}
In co-moving coordinates, one can take
\begin{equation}\label{8}
V^{a}=\frac{1}{\sqrt{f}}\delta^{a}_{0},\quad
q^{a}=\frac{q}{\sqrt{h}}\delta^{a}_{3},\quad
l^{a}=\frac{1}{\sqrt{f}}\delta^{a}_{0}+\frac{1}{\sqrt{h}}\delta^{a}_{3},
\end{equation}
here $q$ is a function of $t$ and $z$.

Using Eq.(\ref{8}), the non-vanishing components of the shear
tensor $\sigma_{ab}$ turn out to be
\begin{equation}\label{9}
\sigma_{11}=-\frac{g}{3}\sigma=\sigma_{22},\quad
\sigma_{33}=\frac{2h}{3}\sigma,
\end{equation}
where
\begin{equation}\label{10}
\sigma=\frac{1}{2\sqrt{f}}\left(\frac{\dot{h}}{h}-\frac{\dot{g}}{g}\right).
\end{equation}
Thus we have
\begin{equation}\label{11}
\sigma_{ab}\sigma^{ab}=\frac{2}{3}\sigma^{2}.
\end{equation}
Also, in view of Eqs.(\ref{3}) and (\ref{4}), it yields
\begin{equation}\label{12}
\pi_{0a}=0, \pi^{3}_{3}= -2\pi^{2}_{2}=-2\pi^{1}_{1}.
\end{equation}
In compact form, it can be written as
\begin{equation}\label{13}
\pi_{ab}=\Omega({\chi}_{a}\chi_{b}-\frac{1}{3}h_{ab}),
\end{equation}
where $\Omega=\frac{3}{2}\pi^{3}_{3}$ and $\chi^{a}$ is a
unit four-vector in $z$-direction satisfying
\begin{equation}\label{14}
\chi^{a}\chi_{a}=1,\quad \chi^{a}V_{a}=0,\quad
\chi^{a}=\frac{1}{\sqrt{h}}\delta^{a}_{3}.
\end{equation}
In view of Eqs.(\ref{6}) and (\ref{8}), it follows that
\begin{equation}\label{15}
a_{3}=\frac{f'}{2f},\quad
\Theta=\frac{1}{\sqrt{f}}\left(\frac{\dot{g}}{g}+\frac{\dot{h}}{2h}\right),
\end{equation}
where dot and prime represent derivative with respect to time $t$ and $z$
respectively.

The Taub's mass for plane symmetric spacetime is defined by
\cite{21}
\begin{equation}\label{16}
m(t,z)=\frac{(g)^{3/2}}{2}R^{12}_{12}=
\frac{1}{8\sqrt{g}}\left(\frac{\dot{g}^2}{f}-\frac{g'^2}{h}\right).
\end{equation}

\section{The Einstein field equations}

The Einstein field equations for the metric (\ref{1}) yield the
following set of equations
\begin{equation}\label{17}
\frac{\dot{g}}{2g}\left(\frac{\dot{g}}{2g}+\frac{\dot{h}}{h}\right)
+\frac{fg'}{2gh}\left(\frac{h'}{h}+\frac{g'}{2g}\right)-\frac{fg''}{gh}
=8\pi(\mu+\epsilon)f,
\end{equation}
\begin{eqnarray}\label{18}
&& \frac{\dot{g}}{2f}\left(\frac{\dot{f}}{2f}+\frac{\dot{g}}{2g}
-\frac{\dot{h}}{2h}\right)
+\frac{g'}{4h}\left(\frac{f'}{f}-\frac{h'}{h}
-\frac{g'}{g}\right)-\frac{f'g}{4fh}\left(\frac{h'}{h}+\frac{f'}{f}\right)\nonumber\\
&+&\frac{\dot{h}g}{4fh}\left(\frac{\dot{h}}{h}+\frac{\dot{f}}{f}\right)
-\frac{\ddot{g}}{2f}+\frac{g''}{2h}+\frac{g}{2fh}(f''-\ddot{h})
=8\pi(p+\Pi-\frac{1}{3}\Omega)g,\nonumber\\
\end{eqnarray}
\begin{equation}\label{19}
\frac{g'}{2g}\left(\frac{g'}{2g}+\frac{f'}{f}\right)+\frac{\dot{g}h}{2fg}\left
(\frac{\dot{g}}{2g}+\frac{\dot{f}}{f}\right)-\frac{\ddot{g}h}{fg}
=8\pi(p+\Pi+\epsilon+\frac{2}{3}\Omega)h,
\end{equation}
\begin{equation}\label{20}
\frac{\dot{g}}{2g}\left(\frac{g'}{g}+\frac{f'}{f}\right)+\frac{g'\dot{h}}{2gh}-
\frac{\dot{g}'}{g}=-8\pi(q+\epsilon)\sqrt{fh}.
\end{equation}
After some manipulation, we can also write Eq.(\ref{20}) in the
following form
\begin{equation}\label{21}
4\pi(q+\epsilon)\sqrt{h}=\frac{1}{3}(\Theta-\sigma)'-\sigma\frac{\sqrt{g}~'}{\sqrt{g}}.
\end{equation}

\section{Junction conditions}

We discuss junction conditions for the interior region $V^{-}$ given
by Eq.(\ref{1}) and the exterior region $V^+$ which is taken as
plane symmetric Vaidya spacetime ansatz given by the line element
\cite{22}
\begin{equation}\label{22}
ds^{2}_{+}=\frac{2m(\nu)}{Z}d\nu^{2}-2d{\nu}dZ+Z^{2}(dX^{2}+dY^{2}),
\end{equation}
where $\chi^{+\mu}\equiv \{\nu,X,Y,Z\}~(\mu=0,1,2,3)$, $\nu$ is the
retarded time and $m(\nu)$ represents total mass inside $\Sigma$.
The line element for the hypersurface $\Sigma$ is defined as
\begin{equation}\label{23}
(ds^{2})_{\Sigma}=-{d{\tau}}^2+A^{2}(\tau)(dx^{2}+dy^{2}),
\end{equation}
where $\xi^{i}\equiv(\tau,x,y)~(i=0,1,2)$ are the intrinsic
coordinates of $\Sigma$.

The Darmois junction conditions \cite{4} are
\begin{itemize}
\item The continuity of the line elements over the hypersurface $\Sigma$ gives
\begin{equation}\label{24}
(ds^{2})_{\Sigma}=(ds^{2}_{-})_{\Sigma}=(ds^{2}_{+})_{\Sigma}.
\end{equation}
This is called continuity of the first
fundamental form.
\item  The continuity of the extrinsic curvature $K_{ab}$ over the hypersurface $\Sigma$
yields
\begin{equation}\label{25}
[K_{ij}]=K^{+}_{ij}-K^{-}_{ij}=0,\quad(a,b=0,1,2).
\end{equation}
This is known as continuity of the second fundamental form.
\end{itemize}Here $K^{\pm}_{ij}$ is the extrinsic curvature defined as
\begin{equation}\label{26}
K^{\pm}_{ij}=-n^{\pm}_{\sigma}(\frac{{\partial}^2\chi^{\sigma}_{\pm}}
{{\partial}{\xi}^i{\partial}{\xi}^j}+{\Gamma}^{\sigma}_{{\mu}{\nu}}
\frac{{{\partial}\chi^{\mu}_{\pm}}{{\partial}\chi^{\nu}_{\pm}}}
{{\partial}{\xi}^i{\partial}{\xi}^j}),\quad({\sigma},
{\mu},{\nu}=0,1,2,3).
\end{equation}
where $n^{\pm}_{\sigma}$ are the components of outward unit normal
to hypersurface $\Sigma$ in the coordinates $\chi^{{\pm}\mu}$.

The equations of hypersurface $\Sigma$ in terms of coordinates $\chi^{{\mp}\mu}$
are given as
\begin{eqnarray}\label{27}
k_{-}(t,z)&=&z-z_{\Sigma}=0,\\
k_{+}(\nu,Z)&=&Z-Z_{\Sigma}(\nu)=0,\label{28}
\end{eqnarray}
where $z_{\Sigma}$ is taken to be an arbitrary constant. Using
Eqs.(\ref{27}) and (\ref{28}), the interior and exterior metrics
take the following form over hypersurface $\Sigma$
\begin{eqnarray}\label{29}
(ds^{2}_{-})_{\Sigma}&=&-f(t,z_{\Sigma})dt^{2}+g(t,z_{\Sigma})(dx^{2}+dy^{2}),\\
\label{30} (ds^{2}_{+})_{\Sigma}&=&2\left(\frac{m(\nu)}
{Z_{\Sigma}}-\frac{dZ_{\Sigma}}{d\nu}\right)d\nu^{2}+Z^{2}_{\Sigma}
(dX^{2}+dY^{2}).
\end{eqnarray}
In view of junction condition (\ref{24}), we get
\begin{eqnarray}\label{31}
Z^{2}_{\Sigma}&=&g(t,z_{\Sigma}),\\ \label{32}
\frac{dt}{d\tau}&=&\frac{1}{\sqrt{f}},\\ \label{33}
\frac{d\nu}{d\tau}&=&\left(2\frac{dZ_{\Sigma}}{d\nu}-\frac{2m(\nu)}{Z_{\Sigma}}\right)^{-1/2}.
\end{eqnarray}
Using Eqs.(\ref{27}) and (\ref{28}), the unit normals in $V^{-}$ and $V^{+}$
respectively, turn out to be
\begin{eqnarray}\label{34}
n^{-}_{\mu}&=&\sqrt{h}(0,0,0,1),\\ \label{35}
n^{+}_{\mu}&=&\left[2\left(\frac{dZ}{d\nu}-\frac{m(\nu)}{Z}\right)\right]
^{-1/2}\left(-\frac{dZ}{d\nu},0,0,1\right).
\end{eqnarray}
The non-zero components of the extrinsic curvature $K^{\pm}_{ij}$ are
\begin{eqnarray}\label{36}
K^{-}_{00}&=&-\left(\frac{f'}{2f\sqrt{h}}\right)_{\Sigma},\\ \label{37}
K^{+}_{00}&=&\left[\frac{d^{2}\nu}{d\tau^{2}}\left(\frac{d\nu}{d\tau}\right)^{-1}
-\frac{m}{Z^{2}}\frac{d\nu}{d\tau}\right]_{\Sigma},\\ \label{38}
K^{-}_{11}&=&K^{-}_{22}=\left(\frac{g'}{2\sqrt{h}}\right)_{\Sigma},\\ \label{39}
K^{+}_{11}&=&K^{+}_{22}=\left[Z\frac{dZ}{d\tau}-2m\frac{d\nu}{d\tau}\right]_{\Sigma}.
\end{eqnarray}

Now, by the junction condition (\ref{25}), i.e., continuity of
extrinsic curvatures, it follows that
\begin{equation}\label{40}
\left[\frac{d^{2}\nu}{d\tau^{2}}\left(\frac{d\nu}{d\tau}\right)^{-1}
-\frac{m}{Z^{2}}\frac{d\nu}{d\tau}\right]_{\Sigma}=
-\left(\frac{f'}{2f\sqrt{h}}\right)_{\Sigma},
\end{equation}
\begin{equation}\label{41}
2m\frac{d\nu}{d\tau}=\frac{\dot{g}}{2\sqrt{f}}-\frac{g'}{2\sqrt{h}}.
\end{equation}
Using Eqs.(\ref{33}) and (\ref{41}), we obtain
\begin{equation}\label{43a}
\left(\frac{d\nu}{d\tau}\right)^{-1}=\frac{1}{\sqrt{g}}
\left[\frac{\dot{g}}{2\sqrt{f}}+\frac{g'}{2\sqrt{h}}\right].
\end{equation}
Inserting Eq.(\ref{43a}) in (\ref{41}), it follows that
\begin{equation}\label{42}
m(\nu)=\frac{1}{8\sqrt{g}}\left(\frac{\dot{g}^2}{f}-\frac{g'^2}{h}\right)
\end{equation}
and hence
\begin{equation}\label{43}
m(t,z)\overset{\Sigma}{=}m(\nu).
\end{equation}
Differentiating Eq.(\ref{43a}) with respect to $\tau$, and making use of  Eqs.(\ref{42})
and (\ref{43a}), we can write Eq.(\ref{40}) as
\begin{eqnarray}\label{43b}
\frac{1}{2\sqrt{fhg}}\left[\frac{-\dot{g}'}{\sqrt{g}}+\frac{g'\dot{h}}{2h\sqrt{g}}
+\frac{f'\dot{g}}{2f\sqrt{g}}+\frac{\sqrt{h}}{\sqrt{f}}\left\{\frac{{-\ddot{g}}}{\sqrt{g}}
+\frac{\dot{g}\dot{f}}{2f\sqrt{g}}+\sqrt{g}\left(\frac{\dot{g}}{2g}\right)^{2}
\nonumber\right.\right.\\\left.\left.+\frac{f}{4g^{3/2}}
\left(\frac{g'}{\sqrt{h}}\right)^{2}+\frac{f'g'}{2h\sqrt{g}}
+\frac{\sqrt{f}}{\sqrt{h}}\left(\frac{g'\dot{g}}{2g^{3/2}}\right)
\right\}\right]\overset{\Sigma}{=}0.
\end{eqnarray}
Comparing Eq.(\ref{43b}) with Eqs.(\ref{19}) and (\ref{20}), it yields
\begin{equation}\label{43c}
p+\Pi+\frac{2}{3}\Omega=q.
\end{equation}

\section{Dynamical equations}

The energy-momentum conservation, $T^{ab}_{;b}=0$, gives
\begin{eqnarray}\label{44}
T^{ab}_{;b}V_{a}&=&\frac{(\dot{\mu}+\dot{\epsilon})}
{\sqrt{f}}+\frac{(q'+\epsilon')}{\sqrt{h}}
+\frac{\dot{g}}{g\sqrt{f}}(\mu+p+\Pi+\epsilon-\frac{1}{3}\Omega)\nonumber\\
&+&\frac{\dot{h}}{2h\sqrt{f}}(p+\Pi+\mu+2\epsilon+\frac{2}{3}\Omega)
+\frac{(fg)'}{fg}\frac{(q+\epsilon)}{\sqrt{h}}=0
\end{eqnarray}
and
\begin{eqnarray}\label{45}
T^{ab}_{;b}\chi_{a}&=&\frac{1}{\sqrt{f}}(\dot{q}+\dot{\epsilon})
+\frac{1}{\sqrt{f}}(q+\epsilon)\frac{(hg\dot{)}}{hg}+\frac{1}{\sqrt{h}}
(p'+\Pi'+\epsilon'+\frac{2}{3}\Omega')\nonumber\\
&&+\frac{f'}{2f\sqrt{h}}(p+\Pi+\mu+2\epsilon
+\frac{2}{3}\Omega)+\frac{g'}{g\sqrt{h}}(\epsilon+\Omega)=0.
\end{eqnarray}
Now we investigate the dynamical properties of the system using the
Misner and Sharp's \cite{2} perspective. For this purpose, we take
the proper time derivative as
\begin{equation}\label{46}
D_{T}=\frac{1}{\sqrt{f}}\frac{\partial}{\partial{t}},
\end{equation}
and the proper derivative in $z$-direction as
\begin{equation}\label{47}
D_{\tilde{Z}}=\frac{1}{\tilde{Z}'}\frac{\partial}{\partial{z}},
\end{equation}
where
\begin{equation}\label{48}
\tilde{Z}=\sqrt{g}.
 \end{equation}

The velocity $U$ of the collapsing fluid can be defined as the
variation of $\tilde{Z}$ with respect to the proper time
\begin{equation}\label{49}
U=D_{T}(\tilde{Z})=\frac{1}{2\sqrt{g}}D_{T}g.
\end{equation}
In the case of collapse, the velocity of the collapsing fluid must
be negative. In view of Eq.(\ref{49}), Eq.(\ref{16}) can take the
following form
\begin{equation}\label{50}
E=\frac{\sqrt{g}~'}{\sqrt{h}}=[U^{2}-\frac{2}{\sqrt{g}}m(t,z)]^{1/2}.
\end{equation}
Making use of Eq.(\ref{47}) in Eq.(\ref{21}), it follows that
\begin{equation}\label{51}
4\pi(q+\epsilon)=E\left[\frac{1}{3}D_{\tilde{Z}}
(\Theta-\sigma)-\frac{\sigma}{\tilde{Z}}\right].
\end{equation}
In case of no dissipation, using Eqs.(\ref{10}), (\ref{15}) and (\ref{49}), the
above equation becomes
\begin{equation}\label{52}
D_{\tilde{Z}}\left(\frac{U}{\tilde{Z}}\right)=0.
\end{equation}
This implies that $U\sim \tilde{Z}$ depicting that now collapse will
be homologous. The rate of change of Taub's mass, using
Eqs.(\ref{16}), (\ref{19}), (\ref{20}) and (\ref{46}), turn out to
be
\begin{equation}\label{53}
D_{T}m=-4{\pi}\tilde{Z}^{2}[(p+\Pi+\epsilon+\frac{2}{3}\Omega)U+(q+\epsilon)E].
\end{equation}
Thus the rate of change of Taub's mass represents variation of total
energy inside the collapsing plane surface. Since this variation is
negative, it shows that total energy is being dissipated during
collapse. The first round brackets on the right hand side stand for
energy due to work being done by the effective isotropic pressure
$(p+\Pi+\frac{2}{3}\Omega)$ and the radiation pressure $\epsilon$.
The second brackets describe energy leaving the system due to heat
flux and radiation. Similarly, using Eqs.(\ref{16}), (\ref{17}),
(\ref{20}) and (\ref{47}), we get
\begin{equation}\label{54}
D_{\tilde{Z}}m=4{\pi}\tilde{Z}^{2}[\mu+\epsilon+(q+\epsilon)\frac{U}{E}].
\end{equation}
This equation describes about the variation of energy between
adjoining plane surfaces inside the fluid distribution. On the right
hand side, $(\mu+\epsilon)$ stands for energy density of the fluid
element plus the energy of null fluid showing dissipation due to
radiation. Moreover, $(q+\epsilon)\frac{U}{E}$ is negative (as
$U<0$), telling that energy is leaving due to outflow of heat and
radiation.

Making use of Eqs.(\ref{16}), (\ref{19}), (\ref{48}) and (\ref{50}),
the acceleration $D_{T}U$ of the collapsing matter inside the
hypersurface $\Sigma$ is given as
\begin{equation}\label{55}
D_{T}U=-4\pi(p+\Pi+\epsilon+\frac{2}{3}\Omega)\tilde{Z}
-\frac{m}{\tilde{Z}^2}+\frac{Ef'}{2f\sqrt{h}}.
\end{equation}
Substituting the value of $\frac{f'}{2f}$ from the above equation
into Eq.(\ref{45}), it follows that
\begin{eqnarray}\label{56}
(p+\Pi+\mu+2\epsilon+\frac{2}{3}\Omega)D_{T}U&=&-(p+\Pi
+\mu+2\epsilon+\frac{2}{3}\Omega)\nonumber\\
&&\times[4\pi{\tilde{Z}}(p+\Pi+\epsilon+\frac{2}{3}\Omega)+\frac{m}{{\tilde{Z}}^2}]\nonumber\\
&&-E^2[D_{\tilde{Z}}(p+\Pi+\epsilon+\frac{2}{3}\Omega)
+\frac{2}{\tilde{Z}}(\epsilon+\Omega)]\nonumber\\
&&-E[D_{T}q+D_{T}\epsilon+4(q+\epsilon)\frac{U}
{\tilde{Z}}+2(q+\epsilon)\sigma].\nonumber\\
\end{eqnarray}
This equation has the form of Newton's second law, i.e.,
\begin{equation*}
Force=Mass\quad density\quad\times\quad Acceleration.
\end{equation*}

The term within the brackets on the left hand side stands for
"effective" inertial mass and the remaining term is acceleration.
The first term on the right hand side represents gravitational
force. Since by the equivalence principle, inertial mass is
equivalent to passive gravitational mass and passive gravitational
mass is equivalent to active gravitational mass. Thus the factor
within round brackets stands for active gravitational mass and the
factor within the square brackets shows how dissipation effects
active gravitational mass. The second square brackets firstly
include gradient of effective pressure which involves radiation
pressure and the collective effect of shear and bulk viscosity. The
second contribution is of local anisotropy of pressure which is the
result of radiation and shear viscosity. The last square brackets
entirely depend upon dissipation. The hydrostatic equilibrium can be
obtained from the above equation by substituting
$U=0,~q=0,~\epsilon=0,~\Pi=0$ and $\Omega=0$.
\begin{equation*}
D_{\tilde{Z}}p=-(\mu+p)\frac{h}{\tilde{Z'}^2}\left[\frac{m}{\tilde{Z}^2}
+4\pi \tilde{Z}p\right].
\end{equation*}

\section{Transport equations}

The general expression for entropy $4$-current is given as \cite{20}
\begin{equation}\label{57}
S^{\mu}=SnV^{\mu}+\frac{q^{\mu}}{T}-(\beta_{0}\Pi^{2}+\beta_{1}q_{\nu}q^{\nu}+
\beta_{2}\pi_{\nu\kappa}\pi^{\nu\kappa})\frac{V^{\mu}}{2T}+\frac{\alpha_{0}\Pi
q^{\mu}}{T}+\frac{\alpha_{1}\pi^{\mu\nu}q_{\nu}}{T},
\end{equation}
where $n$ is particle number density, $T$ is temperature,
$\beta_A(\rho,n)\geq{0}$ are thermodynamic coefficients for
scalar, vector and tensor dissipative contributions to the entropy
density and $\alpha_{A}(\rho,n)$ are thermodynamic viscous/heat
coupling coefficients. The divergence of extended current (follows
from Gibbs equation and Bianchi identities) is given by
\begin{eqnarray}\label{58}
TS^\alpha_{;\alpha}&=&-\Pi\left[V^{\alpha}_{;\alpha}
-\alpha_{0}q^\alpha_{;\alpha}+\beta_{0}\Pi_{;\alpha}
V^{\alpha}+\frac{T}{2}\left(\frac{\beta_{0}}{T}
V^{\alpha}\right)_{;\alpha}\Pi\right]\nonumber\\
&-&q^{\alpha}[h^{\mu}_{\alpha}(\ln{T})_{,\mu}(1
+\alpha_{0}\Pi)+V_{\alpha;\mu}V^{\mu}-\alpha_{0}
\Pi_{;\alpha}-\alpha_{1}\pi^{\mu}_{\alpha;\mu}\nonumber\\
&+&\alpha_{1}\pi^{\mu}_{\alpha}h^{\beta}_{\mu}(\ln{T})_{,\beta}
+\beta_{1}q_{\alpha;\mu}
V^{\mu}+\frac{T}{2}\left(\frac{\beta_{1}}{T}V^{\mu}\right)_{;\mu}
q_\alpha]\nonumber\\
&-&\pi^{\alpha\mu}\left[\sigma_{\alpha\mu}
-\alpha_{1}q_{\mu;\alpha}+\beta_{2}\pi_{\alpha\mu;\nu}V^{\nu}
+\frac{T}{2}\left(\frac{\beta_{2}}{T}V^{\nu}\right)_{;\nu}\pi_{\alpha\mu}\right].
\end{eqnarray}

The $2$nd law of thermodynamics requires that
$S^{\alpha}_{;\alpha}\geq{0}$. This leads to the following transport
equations for our dissipative variables
\begin{equation}\label{59}
{\tau}_{0}\Pi_{,\alpha}V^{\alpha}+\Pi=-\zeta\Theta+\alpha_{0}
{\zeta}q^{\alpha}_{;\alpha}-\frac{1}{2}{\zeta} T
\left(\frac{\tau_{0}}{{\zeta}T}V^\alpha\right)_{;\alpha}\Pi,
\end{equation}
\begin{eqnarray}\label{60}
\tau_{1}h^{\beta}_{\alpha}q_{\beta;\mu}V^{\mu}
+q_{\alpha}&=&-k[h^{\beta}_{\alpha}T_{,\beta}
(1+\alpha_{0}\Pi)+\alpha_{1}\pi^{\mu}_{\alpha}
h^{\beta}_{\mu}T_{,\beta}+T(a_{\alpha}\nonumber\\
&-&\alpha_{0}\Pi_{;\alpha}
-\alpha_{1}\pi^{\mu}_{\alpha;\mu})]-\frac{1}{2}
kT^{2}\left(\frac{\tau_{1}}{kT^2}V^{\beta}\right)_{;\beta}q_{\alpha}
\end{eqnarray}
and
\begin{equation}\label{61}
\tau_{2}h^{\mu}_{\alpha}h^{\nu}_{\beta}\pi_{\mu\nu;\rho}
V^{\rho}+\pi_{\alpha\beta}=-2\eta\sigma_{\alpha\beta}
+2\eta\alpha_{1}q_{<\beta;\alpha>}-{\eta}T\left(\frac{\tau_{2}}
{{2\eta}T}V^{\nu}\right)_{;\nu}\pi_{\alpha\beta},
\end{equation}
where
\begin{equation}\label{62}
q_{<\beta;\alpha>}=h^{\mu}_{\beta}h^{\nu}_{\alpha}\left(\frac{1}{2}
(q_{\mu;\nu}+q_{\nu;\mu})-\frac{1}{3}q_{\sigma;\kappa}
h^{\sigma\kappa}h_{\mu\nu}\right),
\end{equation}
with $k$ as the thermal conductivity. The relaxation times are given
by
\begin{equation}\label{63}
\tau_{0}=\zeta\beta_{0},\quad\tau_{1}=kT\beta_{1},\quad\tau_{2}=2\eta\beta_{2}.
\end{equation}
Notice that if the thermodynamic coupling coefficients are assumed
to be zero, Eqs.(\ref{59})-(\ref{61}) turn to be
Eqs.(2.21)-(2.23) as given in \cite{20}.
The independent components of Eqs.(\ref{59})-(\ref{61}) are
calculated as follows.
\begin{eqnarray}\label{64}
\tau_{0}\dot{\Pi}&=&-\left(\zeta+\frac{\tau_{0}\Pi}{2}\right)\Theta\sqrt{f}
+\alpha_{0}\zeta\frac{\sqrt{f}}{\sqrt{h}}\left[q'+q\left(\frac{f'}{2f}
+\frac{g'}{g}\right)\right]\nonumber\\
&-&\left[\frac{\zeta T}{2}\left(\frac{\tau_{0}}{\zeta
T}\right)^{.}+\sqrt{f}\right]\Pi,
\end{eqnarray}
\begin{eqnarray}\label{65}
\tau_{1}\dot{q}&=&-k\frac{\sqrt{f}}{\sqrt{h}}
\left[T'(1+\alpha_{0}\Pi+\frac{2}{3}\alpha_{1}\Omega)
+T\left\{\frac{f'}{2f}-\alpha_{0}\Pi''\right.\right.\nonumber\\
&-&\left.\left.\alpha_{1} \left(\frac{2}{3}\Omega'+\frac{f'}{3f}\Omega
+\frac{g'}{g}\Omega\right)\right\}\right]\nonumber\\
&-&q\left[\frac{kT^{2}}{2}\left(\frac{\tau_{1}}{kT^{2}}\right)^{.}
+\frac{\tau_{1}}{2}\Theta\sqrt{f}+\sqrt{f}\right],
\end{eqnarray}
\begin{eqnarray}\label{66}
\tau_{2}\dot{\Omega}&=&-2\sqrt{f}\eta\sigma+\eta\alpha_{1}\frac{\sqrt{f}}{\sqrt{h}}
(2q'-\frac{g'}{g}q)\nonumber\\
&-&\left[\eta T\left(\frac{\tau_{2}}{2\eta
T}\right)^{.}\Omega+\frac{\tau_{2}}{2}\Theta\sqrt{f}\Omega+\Omega\sqrt{f}\right].
\end{eqnarray}

Now we discuss the action of dissipation over dynamics of collapsing
object. We couple these transport equations to dynamical equation
(\ref{56}). Using Eq.(\ref{65}) in Eq.(\ref{56}), it follows that
\begin{eqnarray}\label{67}
&&(\mu+p+\Pi+2\epsilon+\frac{2}{3}\Omega)(1-\Lambda)D_{T}U=
(1-\Lambda)F_{grav}\nonumber+F_{hyd}\\&&+\frac{kE^2}{\tau_1}\left[
D_{\tilde{Z}}T(1+\alpha_{0}\Pi+\frac{2}{3}\alpha_{1}\Omega)
-T\left\{\alpha_{0}D_{\tilde{Z}}\Pi+\frac{2}{3}\alpha_{1}\left(
D_{\tilde{Z}}\Omega\nonumber+\frac{3}{\tilde{Z}}\Omega\right)
\right\}\right]\\&&+E\left[\frac{kT^2q}{2\tau_1}\nonumber
D_{T}\left(\frac{\tau_{1}}{kT^2}\right)-D_{T}
\epsilon\right]-E\left[\left(\frac{3q}{2}+2\epsilon\right)
\Theta-\frac{q}{\tau_1}-2(q+\epsilon)\frac{U}{\tilde{Z}}\right],\\
\end{eqnarray}
where $F_{grav}$ and $F_{hyd}$ are given by
\begin{eqnarray}\label{68}
F_{grav}&=&-(p+\Pi+\mu+2\epsilon+\frac{2}{3}\Omega)\nonumber\\
&\times&\left[m
+4\pi(p+\Pi+\epsilon+\frac{2}{3}\Omega){\tilde{Z}}^3\right]\frac{1}{\tilde{Z}^{2}},
\end{eqnarray}
\begin{equation}\label{69}
F_{hyd}=-E^2\left[D_{\tilde{Z}}(p+\Pi+\epsilon+\frac{2}{3}\Omega)
+2(\epsilon+\Omega)\frac{1}{\tilde{Z}}\right]
\end{equation}
and
\begin{equation}\label{70}
\Lambda=\frac{kT}{\tau_{1}}\left(p+\Pi+\mu+2\epsilon+\frac{2}{3}\Omega\right)^{-1}
\left(1-\frac{2}{3}\alpha_{1}\Omega\right).
\end{equation}
Inserting Eq.(\ref{64}) in Eq.(\ref{67}), we obtain
\begin{eqnarray}\label{71}
&&(p+\Pi+\mu+2\epsilon+\frac{2}{3}\Omega)(1-\Lambda+\Delta)
D_{T}U=(1-\Lambda+\Delta)F_{grav}+F_{hyd}\nonumber\\
&&+\frac{kE^2}{\tau_1}\left[D_{\tilde{Z}}T\left(1+\alpha_{0}
\Pi+\frac{2}{3}\alpha_{1}\Omega\right)-T\left\{\alpha_{0}
D_{\tilde{Z}}\Pi+\frac{2}{3}\alpha_{1}\left(D_{\tilde{Z}}
\Omega+\frac{3}{\tilde{Z}}\Omega\right)\right\}\right]\nonumber\\
&&-E^2\left(p+\Pi+\mu+2\epsilon+\frac{2}{3}\Omega\right)
\Delta\left(\frac{D_{\tilde{Z}}q}{q}+\frac{2}{\tilde{Z}}\right)\nonumber\\
&&+E\left[\frac{kT^{2}q}{2\tau_1}D_{T}
\left(\frac{\tau_1}{kT^2}\right)-D_{T}\epsilon\right]
+E\left[\frac{q}{\tau_1}+2(q+\epsilon)\frac{U}{\tilde{Z}}\right]\nonumber\\
&&+E\frac{\Delta}{\alpha_{0}\zeta{q}}\left(p+\Pi+\mu+2\epsilon
+\frac{2}{3}\Omega\right)\left[\left\{1+\frac{\zeta{T}}{2}
D_{T}\left(\frac{\tau_{0}}{\zeta{T}}\right)\right\}\Pi
+\tau_{0}D_{T}\Pi\right],\nonumber\\
\end{eqnarray}
where $\Delta$ is given by
\begin{equation}\label{72}
\alpha_{0}\zeta{q}\left(p+\Pi+\mu+2\epsilon+\frac{2}{3}\Omega\right)
^{-1}\left(\frac{3q+4\epsilon}{2\zeta+\tau_{0}\Pi}\right).
\end{equation}
Here we see that $(1-\Lambda+\Delta)$ is the major factor that
appears in the dynamical equation after coupling it with the
transport equations. We would like to mention here that
Eq.(\ref{71}) is the plane symmetric version of Eq.(55) in
\cite{17}.

\section{Summary and Conclusion}

Gravitational collapse in a star is an irreversible phenomenon.
Dynamics (such as transport processes) of such non-equilibrium
objects and connection between their dynamics and thermodynamics are
of extensive significance in order to have a better visualization of
this problem. Thus we have studied the dynamics of dissipative
collapse, i.e., what role does dissipation play with passing time as
star collapses under the influence of its own gravity.  The most
realistic model of matter, i.e, complicated fluid is assumed in the
interior region and is taken to be consistent with plane symmetry.

To see how system evolves with time, dynamical equations for the
plane symmetric spacetime are obtained using Misner and Sharp
formalism. In the dynamical equation (\ref{56}), we see that the
gravitational force represented by the first term on the right hand
side is expected to be much effective as compared to non-dissipative
fluid and so gravitational collapse is expected to be faster in this
case. Moreover, since the pressure gradient is negative in the
second term on right hand side of this equation, which combined with
the minus sign preceding that term makes a positive contribution,
thereby reducing the rate of collapse. The last square brackets
entirely depend on dissipation and one cannot expect any such
contribution in a dynamical equation for non-dissipative collapse.
The third term in this bracket is positive due to negative sign of
velocity of collapsing fluid U. It shows that outflow of heat flux
$q>0$ and radiation $\epsilon>0$ reduces the total energy of the
system and hence reduces the rate of collapse.

Transport equations in the context of M$\ddot{u}$ller, Israel and
Stewart theory of dissipative fluids are obtained and coupled to
dynamical equation in order to see the influence of dissipation over
dynamics of a collapsing plane. After this union of dynamical and
transport equations, we get equation (\ref{71}) where the factor
$(1-\Lambda+\Delta)$ appears in the dynamical equation. We see the
effect of this factor for different possible values.
\begin{itemize}
\item If $0<(\Lambda-\Delta)<1$, inertial and gravitational
mass densities will be reduced.
\item If $(\Lambda-\Delta)$ tends to $1$, inertial mass density tends to zero.
\item If $(\Lambda-\Delta)>1$, gravitational force will become positive
and it will lead to the reversal of collapse. Another possibility
for reversal of collapse is to take $(\Lambda-\Delta)<1$ such that
$(1-\Lambda+\Delta)$ is sufficiently small. Consequently, it will
significantly decrease the gravitational force.
\end{itemize}


\begin{thebibliography}{40}

\bibitem{1} Oppenheimer, J.R. and Snyder, H.: Phys. Rev. \textbf{56}(1939)455.

\bibitem{2} Misner, C.W. and Sharp, D.: Phys. Rev. \textbf{136}(1964)B571.

\bibitem{3} Vaidya, P.C.: Proc. Indian Acad. Sci. \textbf{A33}(1951)264.

\bibitem{4} Darmois, G.: Memorial des Sciences Mathematiques (Gautheir-Villars, Paris, 1927) Fasc. 25.

\bibitem{5} Sharif, M. and Ahmad, Z.: Mod. Phys. Lett. \textbf{A22}(2007)1493; \textit{ibid.}
2947.

\bibitem{6} Sharif, M. and Ahmad, Z.: Int. J. Mod. Phys. \textbf{A23}(2008)181.

\bibitem{7} Goswami, R.: \emph{Gravitational Collapse of Dustlike Matter with Heat Flux} gr-qc/0707.1122.

\bibitem{8} Nath, S., Debnathm U. and Chakraborty, S.: Astrophys.
Space Sci. \textbf{313}(2008)431.

\bibitem{9} Ghosh, S.G. and Deshkar, D.W.: Int. J. Mod. Phys. \textbf{D12}(2003)317.

\bibitem{10} Chan, R. Mon. Not. R. Astron. Soc. \textbf{316}(2000)588.

\bibitem{11} Herrera, L. and Santos, N.O.: Phys. Rev. {\bf
D70}(2004)084004.

\bibitem{12} Di Prisco, A., Herrera, L., Denmat, G.Le., MacCallum,
M.A.H. and Santos, N.O.: Phys. Rev. {\bf D76}(2007)064017.

\bibitem{13} Herrera, L.: Int. J. Mod. Phys. {\bf D15}(2006)2197.

\bibitem{14} Herrera, L. and Martinez, J.: Astrophys. Space Sci.
{\bf 259}(1998)235.

\bibitem{15} Herrera, L., Denmat, G.Le., Santos, N.O. and Wang, A.:
Int. J. Mod. Phys. {\bf D13}(2004)583.

\bibitem{16} Herrera, L., Di Prisco, A., Hernandez-Pastora, J.L. and
Santos, N.O.: Phys. Lett. \textbf{A237}(1998)113.

\bibitem{17} Herrera, L., Di Prisco, A., Fuenmayor, E. and
Troconis, O.: Int. J. Mod. Phys. \textbf{D18}(2009)129.

\bibitem{18} M$\ddot{u}$ller, I.: Z. Phys. \textbf{198}(1967)329.

\bibitem{19} Israel, W. and Stewart, J.M.: Ann. Phys. (N.Y.)
\textbf{118}(1979)341.

\bibitem{20} Maartens, R.: \emph{Causal Thermodynamics in Relativity}, Lectures
given at the Hanno Rund Workshop on Relativity and Thermodynamics,
(University of Natal, June 1996), astro-ph/9609119.

\bibitem{21} Zannias, T.: Phys. Rev. \textbf{D41}(1990)3252.

\bibitem{22} Debnath, U., Nath, S. and Chakraborty, S.:
Gen. Relativ. Gravit. \textbf{37}(2005)215.












\end{thebibliography}
\end{document}